\documentclass[pra,twocolumn,showpacs,superscriptaddress,amsfonts,amsmath,floatfix]{revtex4}

\usepackage{graphicx}

\newcommand{\Journal}[4]{#1 {\bf #2}, #3 (#4)}
\newcommand{\PR}{Phys. Rev.}
\newcommand{\PRL}{Phys. Rev. Lett.}

\newcommand{\JMP}{J. Math. Phys.}

\newcommand{\Science}{Science}
\newcommand{\Nature}{Nature}

\newcommand{\Exp}[1]{\mathrm{e}^{#1}}

\begin{document}

\title {Quantum fluctuations of a Bose-Josephson junction in a quasi-one-dimensional ring trap}
\author{N. Didier}
\email{Nicolas.Didier@grenoble.cnrs.fr}
\affiliation{Universit\'e Joseph Fourier, Laboratoire de Physique et de Mod\'elisation des Milieux Condens\'es, C.N.R.S. B.P. 166, 38042 Grenoble, France}
\author{A. Minguzzi}
\email{Anna.Minguzzi@grenoble.cnrs.fr}
\affiliation{Universit\'e Joseph Fourier, Laboratoire de Physique et de Mod\'elisation des Milieux Condens\'es, C.N.R.S. B.P. 166, 38042 Grenoble, France}
\author{F.W.J. Hekking}
\email{Frank.Hekking@grenoble.cnrs.fr}
\affiliation{Universit\'e Joseph Fourier, Laboratoire de Physique et de Mod\'elisation des Milieux Condens\'es, C.N.R.S. B.P. 166, 38042 Grenoble, France}

\begin{abstract}
Using a  Luttinger-liquid approach we study the quantum fluctuations of a Bose-Josephson
junction, consisting of a Bose gas confined to a quasi one-dimensional ring trap which
contains a localized repulsive potential barrier. For an infinite barrier we study the
one-particle and two-particle static correlation functions. For the one-body
density-matrix we obtain different power-law decays depending on the location of the
probe points with respect to the position of the barrier. This quasi-long range order can
be experimentally probed in principle using an interference measurement. The
corresponding momentum distribution at small momenta is also shown to be affected by the
presence of the barrier and to display the universal power-law behavior expected for an
interacting 1D fluid. We also evaluate the particle density profile, and by comparing
with the exact results in the Tonks-Girardeau limit we fix the nonuniversal parameters of
the Luttinger-liquid theory. Once the parameters are determined from one-body properties,
we evaluate the density-density correlation function, finding a remarkable agreement
between the Luttinger liquid predictions and the exact result in the Tonks-Girardeau
limit, even at the length scale of the Friedel-like oscillations which characterize the
behavior of the density-density correlation function at intermediate distance. Finally, for a large but finite barrier 
we use the one-body correlation function to estimate the effect of quantum fluctuations
on the renormalization of the barrier height, finding a reduction of the effective
Josephson coupling energy, which depends on the length of the ring and on the interaction strength.
\end{abstract}
\maketitle

\section{Introduction}

The possibility to study Bose-Einstein condensates confined to ring traps constitutes one
of the frontiers of the experimental progress with ultracold atomic gases~\cite{exp_rings}.
The nontrivial topology of these traps together with the uniformity of
the potential along the ring circumference makes them an ideal system for investigating
persistent currents and superfluid properties of the gas.

\begin{figure}
\includegraphics[height=4cm]{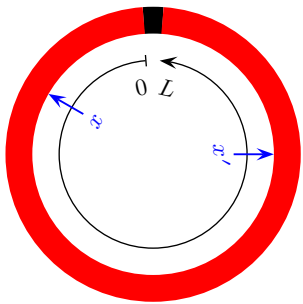}
\caption{Sketch of the Bose-Josephson junction on a ring trap studied in this work.}
\label{fig0}
\end{figure}

While current experimental setups display relatively weak transverse (\textit{\textit{i.e.}}~radial)
confinements, we analyze in this work the case where the transverse confinement is so
strong that only longitudinal (\textit{i.e.}~tangential) quasi-1D motion is allowed along the
ring. In a quasi-1D geometry, the phase coherence properties of the gas are drastically
changed with respect to their 3D counterparts. Phase fluctuations destroy true long-range
order, and by increasing the interaction strength the gas changes from a
quasi-condensate, \textit{i.e.}~a condensate with fluctuating phase~\cite{Popov_book,Petrov00} to
a Tonks-Girardeau gas~\cite{Girardeau60}, where repulsions are so strong that they mimic
the effect of Pauli pressure in a Fermi gas and the condensate is strongly depleted. Here
we consider such a one-dimensional ring trap containing a localized repulsive potential
which creates a ``weak link'' connecting the two ends of the loop~(see Fig.~\ref{fig0}),
a situation that may be viewed as a realization of a Bose-Josephson junction.
Bose-Josephson junctions have been already experimentally realized using a double-well
geometry and arrays~\cite{BJJ_exp}. In the configuration considered here, quantum
fluctuations tend to destroy the phase coherence along the ring, while the tunneling of
bosons between the ends of the loop favors a well-defined phase difference across the
barrier. We will study the interplay between these competing effects.

We start by investigating how the presence of the barrier affects the quantum
fluctuations and hence the coherence properties of a Bose-Josephson junction in a
quasi-1D ring at arbitrary values of the interaction strength. In the absence of the
barrier, these properties have been extensively studied, employing a variety of
techniques, from low-energy Luttinger-liquid approaches and conformal field
theory~\cite{Cazalilla03,Giamarchi_book}, to  exact methods in the Tonks-Girardeau
regime~\cite{Lenard72,Forrester03}. In the presence of an infinitely  high barrier, using the
Luttinger liquid approach for a finite ring, we evaluate here the first-order correlation function, which
describes the decay of phase coherence along the ring, and
recover a previous result from conformal field theory~\cite{Cazalilla03}. We then
compare the results for the one-body density matrix and for the particle-density profile
along the ring with the corresponding exact results in the Tonks-Girardeau limit of impenetrable bosons obtained through a Bose-Fermi mapping
method~\cite{Girardeau60}. This enables us to determine the numerical values of the
nonuniversal parameters of the Luttinger-liquid theory. Knowing these parameters we
estimate two-particle properties such as the density-density correlation function. Finally, we turn to the case of a large but finite barrier, treating the tunneling across the barrier as a perturbation. We
 use the results for the first-order correlation function in the infinite-barrier limit to estimate the effect
of the quantum fluctuations on the effective height of the barrier, \textit{i.e.}~on the
Josephson coupling energy. As a result, we predict how the renormalization of the Josephson  energy depends on the ring length and on the interaction strength.

\section{Luttinger liquid description for a ring with a localized barrier}

We consider $N$ bosons  confined to a uniform, quasi-1D ring-shaped trap of circumference
$L$ to which a localized repulsive potential $V_\mathrm{barr}(x)$, located at $x=0\equiv L$, has
been superimposed. The bosons interact with each other through a repulsive contact potential
$v(x-x')=g\,\delta(x-x')$. The corresponding Hamiltonian in terms
of the bosonic field operators $\Psi(x)$, $\Psi^\dag(x)$ reads
\begin{eqnarray}
\label{eq:ham}
\mathcal{H}&=&\int \mathrm{d}x\,\Psi^\dag(x) \left(\frac{\hbar^2}{2m} \nabla^2 + V_\mathrm{barr}(x)\right)\Psi(x)\nonumber \\
&&+\frac g 2 \int \mathrm{d}x\,\Psi^\dag(x) \Psi^\dag(x)\Psi(x)\Psi(x).
\end{eqnarray}
The presence on the ring of the (very large) barrier potential will be taken into account
below by imposing open  boundary conditions at $x=0=L$ (see Sec.~\ref{sec:obc}) and by
adding a tunnel term (see Sec.~\ref{sec:jos}).

\subsection{Low-energy theory}

In order to evaluate the equilibrium correlation functions at large and intermediate
distances we adopt the Luttinger liquid approach, \textit{i.e.}~we approximate the system
Hamiltonian~(\ref{eq:ham}) by the following effective low-energy Hamiltonian in terms of
the fields $\theta(x)$ and $\phi(x)$ which describe the density and phase fluctuations on
the ring~\cite{Haldane81},
\begin{equation}
\label{eq:hamLL}
\mathcal{H}_{LL}=\frac{\hbar v_s}{2\pi}\int\limits_0^L\mathrm{d}x\left[K\left(\nabla\phi(x)\right)^2+
\frac{1}{K}\left(\nabla\theta(x)\right)^2\right].
\end{equation}
The parameters $K$ and $v_s$ are related to the microscopic interaction parameter of the
original Hamiltonian~(\ref{eq:ham})~\cite{Cazalilla03}, the phase field $\phi(x)$ is
related to the velocity of the fluid $v(x)=\hbar \nabla \phi(x)/m$ and the field
$\theta(x)$ defines the fluctuations in the density profile $\rho(x)$ according to
\begin{equation}
\rho(x)=[\rho_0+\Pi(x)]\sum_{m=-\infty}^{+\infty}\Exp{2mi\theta(x)+i 2\pi  m\rho_0
x+2im\theta_B}, \label{eq:rho1}
\end{equation}
where $\rho_0=N/L$ is the average density of the fluid, $\Pi(x)=\nabla \theta(x)/\pi$ and
$\theta_B$ is a constant fixing the position of the first particle with respect to the
origin of the $x$ axis. The fields $\Pi$ and $\phi$ satisfy the commutation relation~\cite{Haldane81,Cazalilla03}
\begin{equation}
[\Pi(x),\phi(x')]=i \delta(x-x').
\label{eq:com_rel}
\end{equation}
Note that this approach includes not only the lowest-order hydrodynamic expression for
the density fluctuation at long wavelength: the higher order terms in the
sum~(\ref{eq:rho1}) enable the description of the discrete nature of the particles up to
distances $\alpha \sim 1/\rho_0$. Our approach does not allow however to probe shorter
distance scales because of the assumption of linear phonon dispersion modes in the
effective Hamiltonian~(\ref{eq:hamLL}). The bosonic field operator is obtained from the
hydrodynamic expression $\Psi^\dag(x)=\sqrt{\rho(x)} e^{-i\phi(x)}$ and  reads
\begin{eqnarray}
\label{eq:psi}
\Psi^\dag(x)&=&\mathcal{A}\sqrt{\rho_0+\Pi(x)}\nonumber \\
&\times& \sum_{m=-\infty}^{+\infty}\Exp{2mi\theta(x)+2i m
\pi \rho_0 x+2im\theta_B}\,\Exp{-i\phi(x)},
\end{eqnarray}
where~$\mathcal{A}$ is a nonuniversal constant, the value of which depends on the way the
Luttinger liquid approach is regularized in the short wavelength limit. This issue will
be discussed in Sec.~\ref{sec:parameters} below, where the value of the constants
$\mathcal{A}$ and $\theta_B$ will be fixed. The field operators $\Psi$ and
$\Psi^\dagger$ as defined through Eq.~(\ref{eq:psi}) satisfy the standard bosonic commutation relations as a
consequence of the commutation relation~(\ref{eq:com_rel}) among the field operators $\Pi$ and $\phi$.

\subsection{Mode expansion of the Luttinger fields $\mathbf{\theta}$ and $\mathbf{\phi}$ with open boundary conditions}
\label{sec:obc}

We start by considering the case of an infinitely high barrier, which corresponds to a
ring with open boundary conditions. In Sec.~\ref{sec:jos} we will treat the case of a
large, finite barrier by considering the tunneling among the two sides of the barrier as
a perturbation.

In order to evaluate the first- and second-order correlation functions for the bosons on
the ring junction, we derive here the expansion of the fields  $\Pi(x)=\nabla
\theta(x)/\pi$ and $\phi(x)$ in terms of canonical bosonic annihilation and creation
operators~$b_k$ and~$b_k^\dag$ satisfying the commutation relations
$[b_k,b_{k'}^\dag]=\delta_{k,k'}$. Specifically, we expand the operators $\phi(x)$ and
$\Pi(x)$ in Fourier modes for $x\in[0,L]$:
\begin{eqnarray}
\phi(x)&=&\sum_{j=-\infty}^{+\infty}
[\phi_{1,j}b_{k_j}+\phi_{2,j}b_{k_j}^\dag] \,\Exp{i k_j x},\\
\Pi(x)&=&\sum_{j=-\infty}^{+\infty} [\Pi_{1,j}b_{k_j}+\Pi_{2,j}b_{k_j}^\dag]
\,\Exp{i k_j x},
\end{eqnarray}
where we have set $k_j=\frac{2\pi}{L}p j$. The constant $p$ and the complex coefficients
 $\phi_{1,j}$, $\phi_{2,j}$, $\Pi_{1,j}$ and $\Pi_{2,j}$ are determined by imposing three constraints:
(i)  the commutations rules $[\Pi(x),\phi(x')]=i\,\delta(x-x')$; (ii) the open boundary
conditions, which imply vanishing current density  \textit{i.e.}~$\nabla\phi(0)=\nabla\phi(L)=0$,
$\nabla\Pi(0)=\nabla\Pi(L)=0$; (iii) reduction of the Hamiltonian to the diagonal form
$\mathcal{H}=\sum_{k_j}\hbar\omega_{k_j}\left[b_{k_j}^\dag b_{k_j}+\frac{1}{2}\right]$.
In order to take into account that we are using an approximate, long wavelength theory,
we introduce a short distance cutoff~$\alpha\sim{\rho_0}^{-1}$ in the sum over the modes.
The final result reads
\begin{eqnarray}
\label{eq:modes} \phi(x)&=&\phi_0 \nonumber\\
&+&\frac{1}{\sqrt{K}}\sum_{j=1}^\infty\frac{1}{\sqrt{j}}\cos(\pi
j x/L)\,\Exp{-\frac{\pi j \alpha}{2L}}\left[b_{k_j}+b_{k_j}^\dag\right],\\
\Pi(x)&=&\Pi_0\nonumber\\
&+&i\frac{\sqrt{K}}{L}\sum_{j=1}^\infty\sqrt{j}\cos(\pi j x /L)\,\Exp{-\frac{\pi j
\alpha}{2L}}\left[b_{k_j}-b_{k_j}^\dag\right],
\end{eqnarray}
the latter implying
\begin{eqnarray}
\label{eq:theta}
\theta(x)&=&\pi\Pi_0x \nonumber\\
&+&i\sqrt{K}\sum_{j=1}^\infty\frac{1}{\sqrt{j}}\sin(\pi j x/L)\Exp{-\frac{\pi j
\alpha}{2L}}\left[b_{k_j}-b_{k_j}^\dag\right],
\end{eqnarray}
where $k_j=\pi j/L$ ($p=1/2$). The zero mode $\Pi_0$ is directly related to the particle
number operator through normalization: $\Pi_0=(N-N_0)/L$ where $N_0=\langle N
\rangle=\rho_0 L$. It is conjugate to the zero-mode phase operator $\phi_0$ such that
$[\Pi_0,\phi_0]=i$. Using this fact, one can explicitly check the commutation rule
between~$\theta(x)$ and~$\phi(x')$ from the mode expansions~(\ref{eq:modes})
and~(\ref{eq:theta}); it turns out to be $\left[\theta(x),\phi(x')\right]=i\pi
\mathrm{u}(x-x')$, where~$\mathrm{u}$ is the unit step function, consistent
with~(\ref{eq:com_rel}). Finally, inspection of the diagonalized form of the Hamiltonian
yields the linear dispersion relation $\omega_{k_j}=\hbar v_S k_j$ for the modes.

\section{Exact description in the Tonks-Girardeau limit $\mathbf{K=1}$}

In the limit of infinitely strong repulsion  between the bosons, which corresponds to the
value $K=1$ for the Luttinger liquid parameter, an exact solution exists for the bosonic
many-body wavefunction $\Phi(x_1,...x_N)$ (in first quantization). We shall use it
throughout this paper in order to test the results of the Luttinger-liquid theory in the
limit $K=1$, thereby fixing the values of its nonuniversal parameters.

The solution, due to Girardeau~\cite{Girardeau60}, is obtained by mapping the bosons onto
a gas of noninteracting, spin-polarized fermions subject to the same external potential.
The bosonic many-body wavefunction $ \Phi(x_1,...x_N)$ is then obtained in terms of the
fermionic one  as
\begin{equation}
\label{eq:psiTG}
\Phi(x_1,...x_N)=A(x_1,...x_N) \Phi_F(x_1...x_N),
\end{equation}
where the mapping function $A(x_1,...x_N)= \Pi_{1\le j\le \ell\le N} {\rm
sign}(x_j-x_\ell)$ ensures the proper symmetry under exchange of two bosons. The
fermionic wavefunction is given by $\Phi_F(x_1,...x_N)=
(1/\sqrt{N!})\det[\psi_j(x_k)]_{j,k=1...N}$, $\psi_j(x)$ being the single particle
orbitals for the given external potential. Note that $\Phi_F$ vanishes every time two
particles meet as required by Pauli's principle, and hence describes well the
impenetrability condition $g\to \infty$ for the bosons. In our specific case, the
orbitals for a ring of circumference $L$ and open boundary conditions are
\begin{equation}
\label{eq:sporbitals} \psi_j(x)=\sqrt{(2/L)}\sin(\pi j x/L)
\end{equation}
with $j=1,...,\infty$.

As a consequence of the Bose-Fermi mapping, all the bosonic properties which do not
depend on the sign of the many-body wavefunction coincide with the corresponding ones of
the mapped Fermi gas. This is the case \textit{e.g.} for the particle density profile and for the
density-density correlation function. Other properties like the one-body density matrix
and the momentum distribution are instead markedly different for bosons as compared to
fermions. In particular, the calculation of the one-body density matrix requires in
principle the calculation of a (N-1)-dimensional integral, which is known to simplify in
some cases. Examples are the homogeneous gas with periodic boundary conditions~\cite{Lenard72}
or the case of a harmonic confinement~\cite{Forrester03}.

\section{One-body density matrix  and momentum distribution in the infinite-barrier limit}
\label{sec:onebody}

In this section we focus on the one-body density matrix and on the momentum distribution
for the case of a bosonic ring of circumference $L$ with an infinite barrier, with the
aim of analyzing the differences with respect to the case of an infinite system, as well
as to the case of a ring in the absence of the barrier.

\subsection{Contribution from phase fluctuations to the one-body density matrix}
\label{sec:phasephase}

The one-body density matrix, defined as
$\mathcal{G}(x,x')=\left\langle\Psi^\dag(x)\Psi(x')\right\rangle$ yields a measure of the
coherence along the ring.  It is possible to measure the one-body density matrix and
off-diagonal long range order experimentally by measuring the interference pattern of atomic matter waves coming from two holes in the trap (see \textit{e.g.}~\cite{Ritter07} for the
case of a cigar-shaped 3D Bose gas). According to Eq.~(\ref{eq:psi}) the bosonic field
operator has three contributions: (i) the phase $\phi(x)$ (ii) the density fluctuation
$\Pi(x)$, and (iii) the higher harmonics of order $2m \theta(x)$ of the density. The most
important contribution to the one-body density matrix at large distances is the one due
to the phase fluctuations which correspond to the lowest-energy modes of the bosonic
fluid in the ring (see \textit{e.g.}~\cite{Popov_book,Giamarchi_book}), while the two latter
contributions give rise to subleading corrections which we do not analyze further here.

To lowest order we approximate the bosonic field operator~(\ref{eq:psi}) as
$\Psi(x)\simeq {\cal A} \sqrt{\rho_0}\,e^{i\phi(x)}$; the problem then reduces to the
computation of the quantum average $\mathcal{G}_0(x,x')={\cal A}^2 \rho_0 \langle
e^{-i\phi(x)} e^{i\phi(x')} \rangle$. Since the Luttinger-liquid Hamiltonian~(\ref{eq:hamLL})
is quadratic in the  field $\phi(x)$ we immediately obtain
$\mathcal{G}_0(x,x')={\cal A}^2
\rho_0\exp\!\left(-\frac{1}{2}\left\langle\left[\phi(x)-\phi(x')\right]^2\right\rangle\right)$.
The phase-phase correlation function is evaluated with the help of the mode expansion~(\ref{eq:modes});
using the fact that the ground-state average over the bosonic modes is
$\langle (b_k+b_k^\dag)(b_l+b_l^\dag)\rangle=\delta_{kl}$ and the property
$\sum_{j=1}^\infty\frac{1}{j}\,\Exp{-\alpha j}\cos(\gamma j)
=-\frac{1}{2}\ln\!\left[1-2\cos\gamma\,\Exp{-\alpha}+\Exp{-2\alpha}\right]$ one readily
obtains
\begin{eqnarray}
\langle\phi(x)\phi(x')\rangle\!\!\!&=&\!\!\!-\frac{1}{4K}\ln\!\left[(\pi/L)^4\left(\alpha^2+d^2(x-x'|2L)\right)\right.\nonumber \\
\!\!\!&&\!\!\!
\phantom{-\frac{1}{4K}\ln\big[}\times\left.\left(\alpha^2+d^2(x+x'|2L)\right)\right],
\end{eqnarray}
$d(x|L)=L|\sin(\pi x/L)|/\pi$ being the cord function. This leads to
\begin{eqnarray}
&&\mathcal{G}_0(x,x')=\rho_0b_{0,0}\nonumber\\
&\times&\left[\frac{\rho_0^{-2}\sqrt{\left[\alpha^2+d^2(2x|2L)\right]\left[\alpha^2+d^2(2x'|2L)\right]}}
{\left[\alpha^2+d^2(x-x'|2L)\right]\left[\alpha^2+d^2(x+x'|2L)\right]}\right]^\frac{1}{4K},
\label{G0}
\end{eqnarray}
where we have introduced the nonuniversal constant $b_{0,0}=
|\mathcal{A}|^2(\rho_0\alpha)^\frac{1}{2K}$. The above expression~(\ref{G0}) yields the
leading-order term for the one-body density matrix at large distances. By taking the
limit $\alpha\to 0$ we recover the result obtained in~\cite{Cazalilla03} using the
methods of conformal field theory.

\begin{figure}
\centering
\includegraphics[height=4cm]{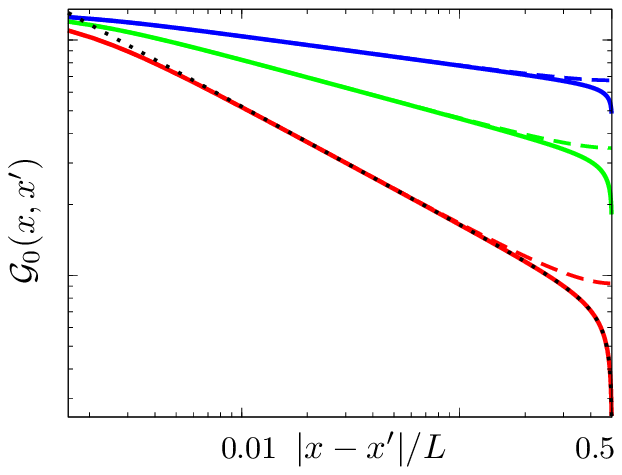}
\includegraphics[height=4cm]{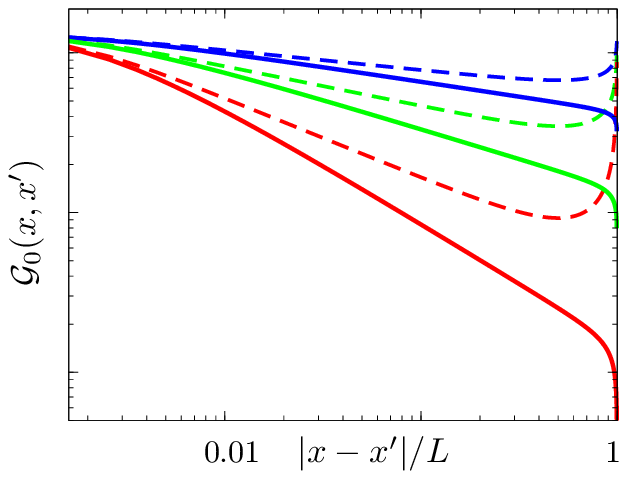}
\includegraphics[height=4cm]{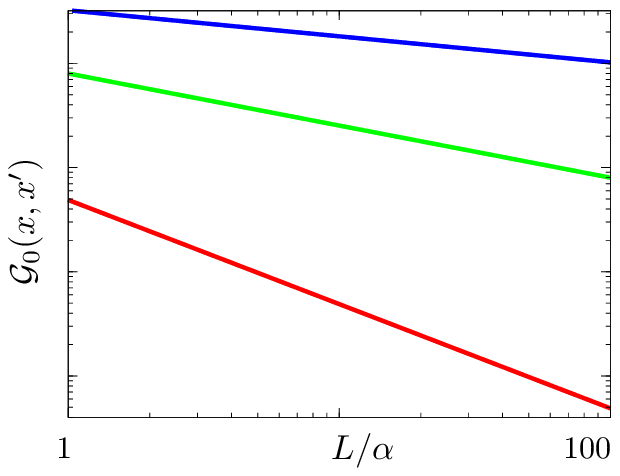}
\caption{(Color online) Top and middle panels: one-body density matrix
  in logarithmic
scale (arbitrary units) as a function of the coordinate $x'$ (in units
of the ring
circumference $L$) for various choices of the coordinate $x$ with
respect to the position
of the barrier, located at $x=0$: top panel $x\simeq0$, middle panel
$x=L/2$. Bottom
panel: one-body density matrix in logarithmic scale (arbitrary
units) taken at
$x\simeq0$ and $x'\simeq L$ as a function of the length $L$ of the
ring (in units of the
average interparticle distance $\alpha$). In each panel we plot three
  values of the parameter~$K$ (from bottom to top $K=1$, $K=2$ and
  $K=4$); the solid lines correspond to the results from
  Eq.~(\ref{G0}) and the dashed lines are the solution~(\ref{G0clean})
  for a homogeneous ring (periodic boundary conditions).
The dotted line in the top panel is the exact solution for a
  Tonks-Girardeau gas in the thermodynamic limit.
The linear behavior of~$\mathcal{G}_0$ in logarithmic scale 
correspond to the predicted
  power law decays with various exponents $\gamma$ (top panel,
  $\gamma=1/2K$, middle panel $\gamma=3/4K$ and bottom panel
  $\gamma=1/K)$. } \label{fig1}

\end{figure}

If the distance among $x$ and~$x'$ is large compared to the cutoff length $\alpha$, the
one-body density matrix displays a power-law decay of the
form~$\mathcal{G}_0(x,x')\propto|x-x'|^{-\gamma}$, where the exponent $\gamma$  can be
derived from the expression~(\ref{G0}), and in particular depends on the location of the
probed points~\cite{Cazalilla03}. Indeed, if the two points are away from the edges one
finds~$\gamma=\frac{1}{2K}$, which corresponds to the result obtained in the
thermodynamic limit~\cite{Haldane81} whereas if they approach the edges
(\textit{i.e.}~$x\alt \alpha$ and~$L-x' \alt \alpha$) the exponent
is~$\gamma=\frac{1}{K}$, a result known in the context of quantum phase fluctuations in a 1D superconducting
wire of length $L$~\cite{Hekking97}. In
the case where one point is at one edge and the other in the bulk we obtain
$\gamma=\frac{3}{4K}$. These three different behaviors are illustrated in
Fig.~\ref{fig1}, where we plot the one-body density matrix $\mathcal{G}_0(x,x')$ as a
function of $x'$ for various choices of the probe point $x$. In the same figure we
display also the behavior for a homogeneous ring in absence of the barrier, obtained by a
procedure analogous to the one outlined above,
\begin{equation}
\mathcal{G}_0^\mathrm{hom}(x,x')=\rho_0b_{0,0}\left[\frac{\rho_0^{-2}}{\alpha^2+d^2(x-x'|L)}\right]^{\frac{1}{4K}}.
\label{G0clean}
\end{equation}
Note that, as the coordinate $x'$ runs along the ring, in the presence of the barrier the
coherence decreases monotonically, while if the barrier is absent coherence is recovered
as $x'$ approaches $L-x$.

The different power-law behaviors are in principle observable for a quasi-1D Bose gas in
a ring trap geometry; it is required to have a high barrier well localized on a length
scale $\alpha$.

\subsection{Momentum distribution}

\begin{figure}
\centering
\includegraphics[height=4.5cm]{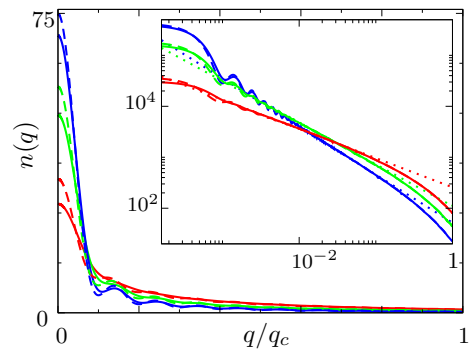}
\caption{(Color online) Momentum distribution $n(q)$ in 
 units of $2\pi|\mathcal{A}|^2\alpha$ as a function of the wavevector $q$ in units of the
 cutoff momentum $q_c=1/\alpha$ for $N=10$ bosons on a ring with
 an infinite barrier (solid lines) and for a homogeneous ring in absence of the
 barrier  (dashed
 lines) for various values of the Luttinger parameter $K$ (from top to bottom
 $K=4,\ 2,\ 1$). The inset shows the same quantity (in logarithmic
 scale, arbitrary units, same line conventions as the main figure) evaluated   for $N=10^3$
 bosons. The dotted lines indicate  the predicted power law decays $q^{1/(2K)-1}$.}
\label{fig:nq}
\end{figure}

We proceed by studying the momentum distribution $n(q)$, obtained by Fourier
transformation of the one-body density matrix with respect to the relative variable,
\begin{equation}
n(q)=\int_0^L\mathrm{d}x\int_0^L\mathrm{d}x'\,\mathrm{e}^{-iq(x-x')}\mathcal{G}(x,x').
\end{equation}
We have resorted to a numerical calculation for the evaluation of the momentum
distribution taking as input the one-body density matrix obtained in Eq.~(\ref{G0}). This
allows to estimate the main features of the momentum distribution at wavevectors $q$
smaller than the cutoff wavevector $q_c\sim 1/\alpha$. The behavior at large wavevectors
$q\gg q_c$ needs an accurate treatment of the short-distance behavior of the many-body
wavefunction~\cite{Minguzzi2002,Olshanii2003}
 and is beyond
the regime of validity  of the Luttinger-liquid method. The result for the momentum
distribution is illustrated in Fig.~\ref{fig:nq} for two values of the boson number in
the ring, and at varying interaction strength. As a general feature (see the inset of
Fig.~\ref{fig:nq}), we observe that at intermediate values of $q$ the momentum
distribution displays a power-law behavior $n(q)\sim q^{1/(2K)-1}$ with the same power
predicted for a  homogeneous ring (see \textit{e.g.}~\cite{Giamarchi_book}). This result
is readily understood as the different power laws described in Sec.~\ref{sec:phasephase}
only occur at the edge of the integration region with a negligible weight with respect to
the bulk contribution. Still, by comparing the details of the momentum distribution of
the ring with the barrier with the momentum distribution of a uniform ring,  (see the
main panel of Fig.~\ref{fig:nq}), we find that in the presence of the barrier the
momentum distribution is decreased at small momenta. This is in agreement with the fact
that the barrier reduces the coherence along the ring. The result is reminiscent of the
one obtained for a 1D gas in presence of disorder~\cite{DeMartino05}, where the reduction
of the momentum distribution at small momenta is also observed.

\section{Particle-density profile and density-density correlation function in the infinite-barrier limit}
\label{sec:rhorho}

Extending the quantum average techniques outlined in Sec.~\ref{sec:onebody} to the limit
of an infinitely high barrier it is possible to evaluate also the inhomogeneous
particle-density profile and the density-density correlation functions. Interference
between particles incident on and reflected by the barrier leads to the occurrence of
Friedel-like oscillations in the density profile and in its correlator, which are typical of
strongly correlated 1D fluids.  We describe here these oscillations within the Luttinger
liquid approach, for any value of the coupling strength, finding that they are more and
more marked as the coupling strength increases. In the Tonks-Girardeau limit of infinite
boson-boson repulsion we compare the predictions of the Luttinger-liquid approach with
the exact results, which enables us to fix the nonuniversal parameters of the latter.

\subsection{Friedel oscillations in the particle density profile}

\label{sec:parameters}

We compute the particle-density profile by taking the quantum average
$\langle\rho(x)\rangle$ of the density operator~(\ref{eq:rho1}) on the ground state,
namely
\begin{equation}
\langle\rho(x)\rangle/\rho_0=\sum_{m=-\infty}^{+\infty}\langle (1+\Pi(x)/\rho_0)  \Exp{2mi\theta(x)}\rangle
\Exp{i 2\pi  m\rho_0 x+2im\theta_B}.
\end{equation}
To evaluate the quantum averages we exploit the fact that  the Hamiltonian~(\ref{eq:hamLL})
is quadratic in the field $\theta(x)$ (and recall that $\Pi(x)=\nabla
\theta(x)/\pi$). We use the property that if $X$ and $Z$ are Gaussian variables then
$\left\langle X\,\Exp{Z}\right\rangle=\langle XZ\rangle\,\Exp{\frac{1}{2}\langle
Z^2\rangle}$. Hence, we are left with the evaluation of the $\langle \theta(x) \theta(x')
\rangle$ correlation function, to be taken at equal points $x=x'$; the  $\langle \Pi(x)
\theta(x') \rangle$ correlation function is obtained from the previous one by taking the
derivative with respect to the variable $x$. Using the mode expansion~(\ref{eq:theta})
and a procedure similar to the one outlined in Sec.~\ref{sec:onebody} we obtain
\begin{eqnarray}
\label{eq:thetatheta}
\langle\theta(x)\theta(x')\rangle&=&\pi^2\langle\Pi_0^2\rangle xx'\nonumber\\
&-&q\frac{K}{4}\ln\!\left[\frac{\alpha^2+d^2(x-x'|2L)}{\alpha^2+d^2(x+x'|2L)}\right],
\end{eqnarray}
and
\begin{eqnarray}
\label{eq:pitheta}
\langle\Pi(x)\theta(x')\rangle\!\!&=&\!\!
\pi\langle\Pi_0^2\rangle x' + \frac{K}{2\pi}\frac{d(x+x'|L)}{\alpha^2+d^2(x+x'|2L)}\nonumber\\
\!\!&-&\!\!\frac{K}{2\pi}\frac{d(x-x'|L)}{\alpha^2+d^2(x-x'|2L)} \mathrm{sign}(x-x').
\end{eqnarray}

The final expression for the density profile reads
\begin{eqnarray}
\label{eq:rho}
\frac{\langle\rho(x)\rangle}{\rho_0}\!\!\!&=&\!\!\!1+2\sum_{m=1}^\infty \left(\frac{\alpha}{d(2x|2L)}\right)^{m^2K} \!\!\!\left[\cos(2m\pi \rho_0  x
+2m\theta_B)\right.\nonumber \\ \!\!\!&-&\!\!\! \left.\frac{m K}{\pi\rho_0} \sin(2m\pi\rho_0 x +2m\theta_B) \frac{d(2x|L)}{d^2(2x|2L)}\right].
\end{eqnarray}
The density profile is modulated by oscillations with wavevector multiples of $2 \pi
\rho_0$. Notice that in the case $K=1$, where the system can be mapped onto a
noninteracting spin-polarized Fermi gas, the wavevectors of the oscillation are multiples
of  $2 k_F$, where $k_F=\pi \rho_0$ is the Fermi wavevector, and hence correspond to the
well-known Friedel oscillations~\cite{Friedel}. For the case of generic $K$ the $m=1$
oscillations decay with the power law~$x^{-K}$ (see \textit{e.g.}~\cite{Egger96}).

Let us now concentrate on the case $K=1$. In the thermodynamic limit ($L\to\infty$,
$N\to\infty$, at fixed $\rho_0=N/L$) the expression~(\ref{eq:rho}) for the density
profile at short distances (to  $O(1/x)$) reduces to
\begin{equation}
\label{eq:rho_LL_TL}
\frac{\langle\rho(x)\rangle}{\rho_0}\simeq 1+\frac{\alpha \cos(2 \pi \rho_0 x + 2 \theta_B)}{x}.
\end{equation}
This can be compared with the thermodynamic limit of the exact expression
 derived using the Bose-Fermi mapping~\cite{Forrester03,Girardeau_impurity}
\begin{equation}
\label{eq:rho_TG_TL}
\frac{\langle\rho(x)\rangle}{\rho_0}\simeq 1-\frac{\sin(2 \pi \rho_0 x)}{2 \pi \rho_0 x},
\end{equation}
allowing us to fix the coefficients $\alpha$ and $\theta_B$ to the values
$\alpha=1/(2\pi\rho_0)$ and $\theta_B=\pi/4$. Note that the latter choice for $\theta_B$
is in agreement with the condition $\theta_B\neq 0, \pm \pi, \pm 2 \pi\dots$ obtained by
imposing that the particle density profile should vanish at $x=0$ and $x=L$~\cite{Cazalilla03}.
Once the constants  $\alpha$ and $\theta_B$ are chosen, the constant
$\mathcal{A}$ in Eq.~(\ref{eq:psi}) can be fixed by comparing the expression for the
coefficient $b_{00}$ entering Eq.~(\ref{G0}) for the one-body density matrix with the
exact value $b_{00}^\mathrm{exact}=2^{-1/3}\sqrt{\pi e} A_G^{-6}\sim 0.521$~\cite{Lenard72,Forrester03}
where $A_G=1.282\dots$ is Glaisher's constant. The result is
$|\mathcal{A}|^2=2^{1/6}\pi\mathrm{e}^{1/2}A_G^{-6}\sim1.307$. This value has
been used in plotting Fig.~\ref{fig1}.

\begin{figure}
\centering
\includegraphics[height=4.5cm]{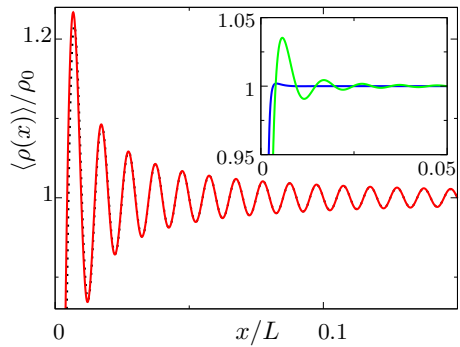}
\caption{(Color online) Particle density profile $\langle\rho(x)\rangle$ in units of the
average particle density $\rho_0$  as a function of the spatial coordinate $x$ along the
ring (in units of the ring circumference $L$) for various  values of the parameter~$K$.
Main figure, $K=1$ (solid line: result from the Luttinger-liquid model, dotted line,
exact result from the Bose-Fermi mapping); inset, $K=2$ (top green line) and $K=4$
(bottom blue line).} \label{figFriedel}
\end{figure}

In Figure~\ref{figFriedel} we illustrate the density profiles for various values of $K$,
obtained by the Luttinger-liquid expression~(\ref{eq:rho}) using the above choice for
$\alpha$ and $\theta_B$~\cite{note}. The figure displays also  the exact result for the
density profile obtained from the Bose-Fermi mapping,
$\rho^\mathrm{exact}(x)=\sum_{j=1}^N |\psi_j(x)|^2$, where the single-particle orbitals
$\psi_j(x)$ are defined in Eq.~(\ref{eq:sporbitals}). The comparison
shows how our choice of parameters $\alpha$ and $\theta_B$ reproduces
extremely well the density profile oscillations even on a finite ring.
The figure also illustrates how the Friedel oscillations display maximal amplitude in the
strongly interacting limit $K=1$.

\subsection{Density-density correlation function from Luttinger-liquid theory}

We turn now to the density-density correlation function $S(x,x')=\langle
\rho(x)\rho(x')\rangle -\langle \rho(x)\rangle \langle \rho(x')\rangle$. This quantity
encodes the information on the structure of the fluid, \textit{i.e.}~on the correlations between
density modulations at different parts of the fluid, while it vanishes for an ideal Bose
gas.  The Fourier transform of the density-density correlation function with respect to
the relative variable is directly accessible experimentally by light-scattering methods
(see \textit{e.g.}~\cite{Vignolo01} and references therein).

The density-density correlation function is obtained with the quantum average method
described in Sec.~\ref{sec:onebody} and~\ref{sec:rhorho}. One has to compute
\begin{eqnarray}
\langle\rho(x)\rho(x')\rangle&=&
\sum_{m,m'=-\infty}^{+\infty}
\mathrm{e}^{2i(m-m')\theta_B}\,\mathrm{e}^{i2\pi\rho_0(mx-m'x')}
\nonumber\\&\times&
\langle\left[{\rho_0}^2+\rho_0(\Pi(x)+\Pi(x'))+\Pi(x)\Pi(x')\right]
\nonumber\\&\times&
\mathrm{e}^{2i(m\theta(x)-m'\theta(x'))}\rangle.
\end{eqnarray}
The average can be performed using the general result for Gaussian variables
$\left\langle XY\mathrm{e}^Z\right\rangle
=\left(\langle XY\rangle+\langle XZ\rangle\langle YZ\rangle\right)\mathrm{e}^{\frac{1}{2}\langle Z^2\rangle}$.
The novel correlator needed for the calculation in addition to Eqs.~(\ref{eq:thetatheta}) and~(\ref{eq:pitheta}) is
\begin{eqnarray}
\langle\Pi(x)\Pi(x')\rangle\!\!\!&=&\!\!\!
-\frac{K}{2\pi^2}\left[\frac{d^2(x-x'|2L)-\alpha^2\cos(\pi(x-x')/L)}{\left(d^2(x-x'|2L)+\alpha^2\right)^2}\right. \nonumber \\\!\!\!&+&\!\!\!\left.\frac{d^2(x+x'|2L)-\alpha^2\cos(\pi(x+x')/L)}{\left(d^2(x+x'|2L)+\alpha^2\right)^2}\right]\nonumber \\\!\!\!&+&\!\!\!\langle\Pi_0^2\rangle.
\end{eqnarray}
The final result reads
\begin{widetext}
\begin{eqnarray}
\langle\rho(x)\rho(x')\rangle&=&
\sum_{m,m'=-\infty}^{+\infty}
(\rho_0\alpha)^{(m^2+m'^2)K}\,\mathrm{e}^{2i(m-m')\theta_B}\mathrm{e}^{i2\pi\rho_0(mx-m'x')}
\nonumber\\&\times&
\Big[\rho_0^2+\langle\Pi(x)\Pi(x')\rangle
+2i\rho_0\langle(\Pi(x)+\Pi(x'))(m\theta(x)-m'\theta(x'))\rangle
\nonumber\\&&\;
-4\langle\Pi(x)(m\theta(x)-m'\theta(x'))\rangle
\langle\Pi(x')(m\theta(x)-m'\theta(x'))\rangle\Big]
\nonumber\\&\times&
\left(\frac{\alpha^2+d^2(x+x'|2L)}{\alpha^2+d^2(x-x'|2L)}\right)^{mm'K}
\left(\frac{\rho_0^{-2}}{\alpha^2+d^2(2x|2L)}\right)^{\frac{1}{2}m^2K}
\left(\frac{\rho_0^{-2}}{\alpha^2+d^2(2x'|2L)}\right)^{\frac{1}{2}m'^2K}.
\end{eqnarray}
\end{widetext}
This equation displays the general structure of the density-density correlations to all
orders in $m$ and $m'$, and by considering only the first terms of the expansion $m,
m'=0,\pm 1$ we recover the known results~\cite{Giamarchi_book,Cazalilla03}.

We proceed by comparing the density-density correlation function $S(x,x')$ with the exact
result for $K=1$.  The latter is obtained from the Bose-Fermi mapping as~\cite{Vignolo01}:
\begin{eqnarray}
S^\mathrm{exact}(x,x')=-\left[\sum_{j=1}^N{\psi_j}^*(x)\psi_j(x')\right]^2,
\label{orbitals}
\end{eqnarray}
 where the single-particle orbitals $\psi_j(x)$ are defined in Eq.~(\ref{eq:sporbitals}).

\begin{figure}
\centering
\includegraphics[height=4.5cm]{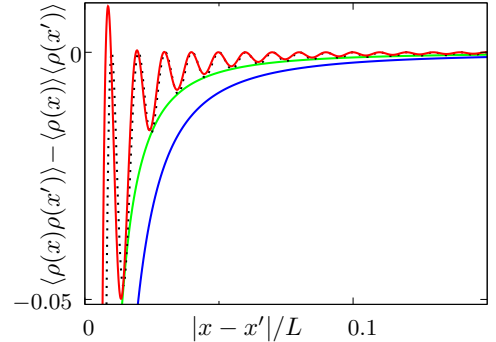}
\caption{(Color online) Density-density  correlation function $S(x,x')$ (in units of
$\rho_0^2$) from the Luttinger-liquid model as a function of the coordinate $x'$ (in
units of the ring circumference $L$), with $x=L/2$ and
  for various values of the Luttinger parameter $K$ ($K=1$, top red line, $K=2$, middle green line, $K=4$ bottom blue
  line).
 The dotted  line corresponds to the exact solution obtained from the Bose-Fermi mapping in the case $K=1$.}
\label{figrhorho}
\end{figure}

Figure~\ref{figrhorho} displays  the results obtained from the Luttinger-liquid method at
various values of the Luttinger parameter $K$, using the choice of parameters  $\alpha$
and $\theta_B$ determined from the density profile in Sec.~\ref{sec:parameters} and
compares to the exact ones in the case $K=1$. The agreement found is very good, even for
the Friedel-like oscillations at wave vector $k\sim 2 \pi \rho_0$; this is at the
boundary of the expected regime of validity of the Luttinger-liquid theory and
illustrates how a reasonable choice of the non-universal parameters in the effective
model allows for surprisingly accurate predictions.

\section{Renormalization of Josephson energy by quantum fluctuations}
\label{sec:jos}

In this section we consider the effect of a finite barrier on the ring located at $x=0\equiv L$, and described by a localized potential $V_{\rm barr}(x)=\mathcal{T} \delta(x)$, where~$\mathcal{T}$ is the tunneling strength of the barrier. This yields 
 the following barrier term in the Hamiltonian~(\ref{eq:ham}):
\begin{equation}
\mathcal{H}_J = \mathcal{T}\,\Psi^\dagger(L) \Psi(0)\,+\,\mathrm{h.c.}\,.
\label{eq:ham_tunnel}
\end{equation}
The above equation takes into account the possibility for a boson to tunnel through the barrier potential, and $h.c.$ refers to the Hermitian conjugate corresponding to tunneling events in the opposite direction.
In the hydrodynamic formulation for~$\Psi^\dag(x)$, the transfer of one boson is ensured by the operator~$\exp(-i\varphi)$ where $\varphi=\phi(L)-\phi(0)$.
By neglecting the density fluctuations in the field operator (\ref{eq:psi}) 
 we then recover the usual Josephson Hamiltonian
\begin{equation}
\mathcal{H}_J = E_J\cos\varphi,
\label{eq:ham_Jos}
\end{equation}
where~$E_J=2\rho_0\mathcal{T}$ is the Josephson energy of the junction.

Quantum fluctuations of the bosons in the ring on both sides of the barrier tend to smear
the phase~$\varphi$ and hence suppress the tunneling strength. Indeed, from the diagonal
Hamiltonian of section~\ref{sec:obc}, the ring constitutes an oscillator bath for the
junction with linear spectrum~$\hbar\omega_{k_j}=\pi\hbar v_S j/L$; the resulting model
is very similar to the one describing a superconducting Josephson junction coupled to a
resistive environment~\cite{Schoen90}. Tunneling events thus induce excitations of the
modes of the ring with energy between~$\hbar\omega_0=\pi\hbar v_S/L$ and the high energy
cutoff~$\hbar\omega_h=\pi\hbar v_S/\alpha$.

When the Josephson energy is smaller than the lowest
mode~$\hbar\omega_0$, corresponding to small rings $L<L^*\sim\hbar
v_S/E_J$, the junction can be treated as a perturbation and every mode
modifies~$E_J$. The effective Josephson energy results from
averaging~$\mathcal{H}_J$, Eq.~(\ref{eq:ham_Jos}), with respect to the
ground state of the unperturbed Hamiltonian~(\ref{eq:hamLL}): $E_J^\mathrm{eff}=\langle \mathcal{H}_J\rangle$ with $\langle \mathcal{H}_J\rangle/E_J=\mathcal{G}_0(L,\alpha)/\rho_0$. Then
\begin{equation}
\label{eq:ej1}
E_J^\mathrm{eff}=E_J\left(\frac{\pi\alpha}{2L}\right)^{1/K}\quad\mathrm{for}\ L<L^*.
\end{equation}
The Josephson energy decreases with the power law~$L^{-1/K}$ of the
one-body density for probe points at the edges of the ring. This case
includes the limit of an infinitely high barrier, where $E_J\to 0$,
$L^*\to \infty$, and which is  illustrated in
Fig.~\ref{fig1}, bottom panel.

\begin{figure}
\centering
\includegraphics[height=4.5cm]{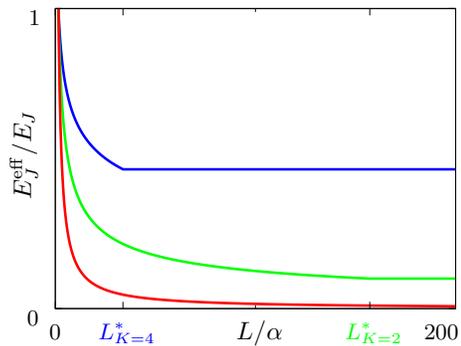}
\caption{(Color online) Renormalized Josephson energy as a function of the length of the ring. From top to bottom $K=4,\ 2,\ 1$ with $\hbar\omega_h=10\,E_J$. $E_J^\mathrm{eff}$ decreases as~$1/L^{1/K}$ and reaches a constant value at~$L=L^*$.}
\label{figEJ}
\end{figure}

When~$E_J$ is larger than $\hbar\omega_0$ (with $E_J<\hbar\omega_h$), only the modes with
energies larger than~$E_J$ contribute to the renormalization. Consequently, to obtain the
effective Josephson energy we need to average over wavelengths between~$\alpha$ and the
characteristic length $\ell\equiv\pi\hbar v_S/E_J^\mathrm{eff}$:
$E_J^\mathrm{eff}=\langle\!\langle \mathcal{H}_J\rangle\!\rangle$ with
\begin{equation}
E_J^\mathrm{eff}/E_J=\mathcal{G}_0(L,\alpha)/\mathcal{G}_0(\ell,\alpha)\sim
(\alpha/\ell)^{1/K}
\end{equation}
The effective Josephson energy in this case is obtained by solving the above
self-consistent equation with respect to~$\ell$, with the result
\begin{equation}
\label{eq:ej2}
E_J^\mathrm{eff}=E_J\left(\frac{\alpha E_J}{\pi\hbar v_S}\right)^{1/(K-1)}\quad\mathrm{for}\ L>L^*.
\end{equation}
In this case $E_J^\mathrm{eff}$ is independent of the ring circumference~$L$. Our
results~(\ref{eq:ej1}) and~(\ref{eq:ej2}) are summarized in  Fig.~\ref{figEJ}. As a main
conclusion, we find  that quantum fluctuations dramatically reduce the tunnel amplitude
with respect to its bare value entering the Hamiltonian~(\ref{eq:ham_tunnel}), especially
in the case~$K=1$ . Note however that the reduction saturates at a nonzero  level for
rings larger than the so-called healing length~$L^*$.
 The
continuity between the two regimes $L<L^*$ and $L> L^*$ defines the healing length of the
ring $L^*=\pi\alpha/2(\pi\hbar v_S/\alpha E_J)^{K/(K-1)}$.

As a final remark, we would like to mention that our approach
is equivalent to the renormalization group formalism~\cite{KaneFisher} or
the self-consistent harmonic approximation~\cite{Schoen90}.

\section{Summary and concluding remarks}

In summary, in this paper we have studied the equilibrium properties of a quasi-1D
interacting Bose gas confined in a ring trap with a localized barrier. In the limit of
infinite barrier we have studied the coherence, density profiles and density-density
correlations of the gas using a Luttinger-liquid approach and the quantum average method.
Our results recover and extend those previously known by the use of conformal field
theory methods. As  physical consequences of our analysis, we find that the one-body
density matrix, when probed at various points with respect to the barrier position, is
expected to display universal power law behaviors with different exponents which depend
only on the Luttinger parameter $K$. We also find that our method permits to describe
accurately the Friedel oscillations (due to the presence of the barrier) occurring in the
particle density profile and in the density-density correlation function. Once the
non-universal parameters entering the effective model are fixed  by comparing the density
profile to the exact one in the Tonks-Girardeau case $K=1$, we find that the Luttinger
liquid model well agrees with the exact result for the density-density correlation
function at a length scale which is at the boundary of the validity of the
Luttinger-liquid model.

The analysis performed in the limit of infinite barrier is then use to study
perturbatively the presence of a large, finite barrier. By taking into account the effect
of quantum fluctuations we find that the effective Josephson energy (\textit{i.e.} the
tunnel amplitude across the barrier) is reduced with respect to its bare value, in a way
which depends on the length of the ring with respect to a typical healing length, a
maximal reduction occurring for long rings.

The effect of the renormalization of the tunnel amplitude is expected to have strong
consequences on the dynamical evolution of the ring-trapped Bose gas with a Josephson
junction.

\begin{acknowledgments}
We acknowledge stimulating discussions with R. Citro, L.I. Glazman, J. Schmiedmayer.
We thank IUF, CNRS and the MIDAS-STREP project for financial support.
\end{acknowledgments}


\begin{thebibliography}{33}
\bibitem{note} In absence of known results for the case of arbitrary $K$, we have adopted the same choice of parameters $\alpha$ and $\theta_B$ for any value of~$K$.
\end{thebibliography}
\end{document}